\documentclass{iopart}             
\usepackage[T1]{fontenc}
\usepackage[applemac]{inputenx}
\usepackage{url,graphics,epsfig}
\usepackage{amssymb}

\begin{document}
\setlength{\arraycolsep}{2.5pt}             
\jl{2}
%
%
%
\def\etal{{\it et al~}}
\def\newblock{\hskip .11em plus .33em minus .07em}
%
%
%
%
%
\title[Photon impact double ionisation of Li$^+$ ions]{An R-matrix with pseudo-state (RMPS) approach to the single photon double ionization 
                                                                and excitation of  the He-like  Li$^{+}$ ion}

\author{B M McLaughlin$^{1,2}$\footnote[1]{Corresponding author, E-mail: b.mclaughlin@qub.ac.uk}}

\address {$^{1}$Centre for Theoretical Atomic, Molecular and Optical Physics (CTAMOP), 
                              School of Mathematics and Physics, Queens University of Belfast,
                              The David Bates Building, 7 College Park, Belfast BT7 1NN, UK}

\address{$^{2}$Institute for Theoretical Atomic and Molecular Physics (ITAMP),
                             Harvard Smithsonian Center for Astrophysics, MS-14,
                             Cambridge, Massachusetts 02138, USA}

%
%

\begin{abstract}
The success of the R-matrix with pseudo-state (RMPS) method  to model 
single photoionization processes, benchmarked against dedicated
synchrotron light source measurements, 
is exploited and extended to investigate,  single
photon double ionization cross-sections for the He-like, Li$^+$ ion.
We investigate these processes from both the ground 
state and the excited $\rm 1s2s^{1,3}S$ metastable levels of this He-like system.
Comparisons  of the results from the R-matrix plus pseudo-state (RMPS) method are made with other 
state-of-the-art theoretical approaches such as time-dependent close-coupling (TDCC), 
B-splines and the convergent close-coupling (CCC).  
Excellent agreement with various other theoretical approaches 
are achieved but differences  occur which are high-lighted and discussed.
For the ground state the peak of the cross section is $\sim$ 2 kilo-barns (Kb), 
that for the $\rm 1s2s^{1}S$ state is $\sim$ 6 Kb and $\sim$ 1 Kb for the corresponding $\rm 1s2s^{3}S$ state.
All the cross sections for single photon double ionization are extremely 
small, being in the region of  2Kb -- 10 Kb, or less, 
rendering their experimental determination extremely challenging.

\end{abstract}
%
%
%

\vspace{0.25cm}
{\begin{flushleft}
Short title: Photon impact double ionization of Li$^{+}$ ions
\\

\vspace{0.25cm}
J. Phys. B: At. Mol. Opt. Phys: \today
\end{flushleft}}
\maketitle
%
%
%
%

\section{Introduction}
Photoionization is one of the most important processes within the interstellar medium (ISM). 
It dominates the heating of the ISM and produces the majority of the ionized medium. 
Photoionization is known to be the excitation mechanism in planetary nebulae, 
H II regions, starburst galaxies and Seyfert narrow line regions. 
Thus modeling photoionization is an important theoretical study as it not 
only can provide a physical insight into these emission line regions 
but can also gain us a better understanding of the ISM as a whole.
The photoionization process is one of the important radiative
feedback processes in astrophysics \cite{Miyake2010,Stancil2010}.  
The increase in pressure caused by it can trigger strong dynamic
effects: photoionization hydrodynamics.  The challenge in combining hydrodynamics with photoionization
lies in the difference in time scales between the two process.

The situation for single photon double photoionization of ions of the He-like
isoelectronic sequence are of importance in plasma
physics and astrophysics, is far less advanced than that of single photoionization.  
This is  due to major difficulties in attaining target densities sufficient to carry out experiments.  
The majority of our knowledge on this topic has  been provided mainly by theoretical studies.
The present study on the single photon double ionisation of Li$^{+}$ ions, a proto-typical 
two-electron system, provides an essential benchmark for future photoionization studies on 
more astrophysically abundant highly ionized He-like species such as C, N, O and Ne  which are 
of great importance, in determining the mass of missing baryons in the X-ray forest of the 
warm-hot intergalactic medium (WHIM) \cite{Nicastro2005,Nicastro2012}.  

We note that X-ray spectra obtained by Chandra, from sources such as Capella, Procyon, and HR 1099, are
used as standards to benchmark plasma spectral modelling codes. He-like ions of O, Ne, Mg, Si, S, and Ar 
were first discovered in the X-ray spectra of the Seyfert 1 galaxy NGC 3783, obtained with the
High Energy Transmission Grating Spectrometer on the Chandra X-Ray Observatory \cite{kaspi2000}.
Strong He-like, carbon [C V] lines have been detected in the Chandra HRC-S/LETG X-ray grating spectrum
of the blazar H 2356-309 \cite{Nicastro2012} at 44.8 \AA.
He-like, nitrogen [N VI] lines have also been observed with Chandra and {\it XMM-Newton} in the X-ray spectra 
of Capella and Procyon \cite{Mewe2001,Ness2001}, the M dwarf binary YY Gem \cite{stelzer2002}, and the recorded outburst of
 the recurrent nova RS Oph \cite{Ness2007},  in the wavelengths region 28.7 \AA  -  29.6 \AA (420 eV)  that
are attributed to 1s $\rightarrow$ 2$\ell$ transitions. 
{\it XMM-Newton} observations of the fast classical nova V2491 Cyg \cite{Ness2011} have also indicated the 
N VI, K$_{\alpha}$  and K$_{\beta}$ lines are present at about 28.78 \AA~ and 24.90 \AA. 
The He-like series lines of N VI and O VII are detected up to 1s  $\rightarrow$ 5p, and
for O VII, the recombination/ionization continuum at 16.77 \AA
(739.3 eV) is present between 16.6 \AA  -  16.8 \AA. 
The intra-cluster medium (ICM) is the few 10$^{6}$  - 10$^{8}$ Kelvin, X-ray emitting plasma 
which fills the potential wells of objects ranging in scale from galaxy clusters down 
to massive elliptical galaxies. He-like oxygen [OVII] has been revealed 
in stacked high spectral resolution, {\it XMM-Newton} Reflection Grating Spectrometer (RGS) 
spectra from galaxy clusters, groups of galaxies and elliptical galaxies \cite{Sanders2011}. 
Therefore, He-like C, N, O, Ne, and Fe are all of interest. The $n$=2
to $n$=1, x-ray lines, are all used as diagnostics of photoionized and
collisionally-ionized gas. Typically, for He-like systems, the line ratios (R, L, and G) are
studied and for double photoionization, high energy
photons are required for applications in Active Galactic Nuclei (AGN's) 
where all elements between, He, Li, Be, B, C, ...., Zn can be treated \cite{Porter2007,Porquet2010}.

As one of the most basic and fundamental three-body Coulomb problems,  
single photon double ionization of  two-electron  (He-like) system requires an accurate
description of the correlated motion of two electrons in the long-range Coulomb field of 
the residual stripped ion. As this process comprises the interaction of just three charged particles, 
it can properly only be handled numerically. With the advancement of computational power 
over the past decade or more, this has opened the doorway to investigate large scale numerical 
studies of this fundamental three-bodied problem using the R-matrix 
with pseudo-states method  \cite{Bartschat1996a,Gorczyca1997a,mit99,Ludlow09,Ludlow09b} 
allowing detailed convergence studies to be made.  We note that the R-matrix with pseudo-states was first 
introduced into the literature, in the late 1960's, by Burke, Gallagher and Geltman\cite{Burke1969} in studying electron 
scattering from atomic hydrogen with great success reproducing the earlier highly accurate work of Schwartz \cite{Schwartz1961}.

Single photon double ionization of the Li$^{+}$ ion has been studied theoretically by Kornberg and Miraglia \cite{kornberg94} using two and three 
screened Coulomb wavefunctions and more recently using the eigenchannel R-matrix method by Meyer \cite{meyer97}. 
Wehlitz and co-workers \cite{wehlitz97} have measured triple photoionization
of lithium and related their experimental triple-to single
photoionization cross sections ratio to the theoretical
double-to-single ratio of the Li$^{+}$ ion reported by Kornberg
and Miraglia \cite{kornberg94}. A two-stage mechanism of triple photoionization 
was proposed. In the first stage double photoionization of the valence 1s$^2$ shell of the
Li atom takes place followed by the
shakeoff of the remaining 2s electron into the continuum.
It is assumed that the double photoionization of
the 1s$^2$ shell in the Li atom and the Li$^{+}$ ion are quite similar
and the resulting triple-to-single photoionization
cross-section ratio for the Li atom can be calculated as the
double-to-single ratio of the Li$^{+}$ ion multiplied by the probability
of the shakeoff ~0.00174 according to Wehlitz et al \cite{wehlitz98}. 
This serves as a useful check on various theoretical models.

Various theoretical methods have been used to explore and study single photon double ionization 
of He-like systems, these are;  the two screened Coulomb method \cite{shake93,shake95,shake96},
the time-dependent close-coupling approach(TDCC) \cite{ulrich05,mitch07,Ludlow09b,mitch10}, 
convergence close-coupling (CCC) \cite{bray98,bray98b,bray00,bray01},  
many-bodied perturbation (MBPT) \cite{hino93,kelly95}, 
B-spline with R-matrix, exterior complex scaling B-spline \cite{hugo2001}, 
Intermediate Energy R-matrix Method (IERM) \cite{Scott2012},
and presently now using the R-matrix with pseudo states (RMPS) 
method \cite{Bartschat1996a,Gorczyca1997a,mit99,Ludlow09,Ludlow09b}. 
It has been shown by Forrey and co-workers \cite{forrey95} that in the limit of
high photon energies double ionization of the excited singlet
metastable state of Li$^{+}$ ions has a larger cross-section ratio than double
ionization of the ground state \cite{Dalgarno1992}.  McCurdy and co-workers \cite{McCurdy2010}
have used a finite-element discrete-variable representation (DVR) and exterior
complex scaling to study excited state single photon double ionisation of Li and Be with excellent 
agreement obtained with results from the R-matrix with pseudo-states (RMPS) approach.  F{\o}rre \cite{forre2012} 
has shown (for single-photon double ionization of helium) a heuristic formula for the cross section
appears to work well compared to ab initio calculations.  Applications of this formulae 
for the double ionisation of Li$^{+}$ compared with detailed ab initio calculations suggests the 
methodology is valid for the entire helium sequence \cite{forre2012}. 

We note in passing for He-like systems,
 Scott and co-workers \cite{Scott2012} recently have shown that the intermediate 
 energy R-matrix (IERM) results for single photon, single and double ionization  
and detachment are in excellent agreement with those obtained from experiment and the RMPS 
method for He \cite{Ludlow2010,Samson1998} and H$^{-}$ \cite{Stancil2010,Miyake2010} giving 
enhanced confidence in the existing theoretical data on H$^{-}$ for astrophysical applications 
in the early universe \cite{Stancil2013}. Projected experimental measurements \cite{Mueller2013b} on this He-like
ion is one of the major motivations for undertaking the present theoretical study.

The single photon double ionization processes investigated here are on the $\rm 1s^2~^1S$ ground state of
the Li$^{+}$ ion are:
\begin{center}
h$\nu$ + Li$^{+}$ (1s$^2$ $^1$S) $\rightarrow$  Li$^{3+}$  + e$^-$+ e$^-$.\\
\end{center}
We extend our investigation  also  to the excited n=2 metastable levels
\begin{center}
h$\nu$ + Li$^{+}$ (1s2s $^{1,3}$S) $\rightarrow$  Li$^{3+}$  + e$^-$+ e$^-$,\\
\end{center}
as the Li$^{+}$ (1s2s $^{1,3}$S) metastable states are of prime interest.
In both cases we have a fully stripped Li nucleus  with the
two out going electrons in the continuum whereas 
in the case  of single photoionization process the resulting H-like Li$^{2+}$ (n$\ell$)
ion may be left in its ground or excited state with one electron in the continuum. 
Figure 1 shows a schematic diagram of the Li$^+$ and Li$^{2+}$ spectrum adapted from the work of 
Kleiman and co-workers  \cite{ulrich05}.
In the previous investigations on this systems we note that for the 
case of single photon single ionization \cite{Scully2006,Scully2007}  
calculations were carried out using the 
R-matrix with pseudo states method (RMPS) \cite{Bartschat1996a,Gorczyca1997a,mit99,Ludlow09,Ludlow09b}.
In that work, the RMPS results were shown to be in excellent agreement with the high resolution measurements made 
at the Advanced Light Source synchrotron radiation facility in Berkeley, California.  Here 
we concentrate our efforts on photo-excitation and double ionization processes for this 
He-like system above the ionization threshold from the ground and metastable 1s2s $^{1,3}$S states. 
We note preliminary investigations for the case of the ground state have been carried out 
by  McLaughlin and Ballance \cite{McLaughlin2009} using this same RMPS method. 
Here we use this same RMPS method and apply it to the case of single photon double ionization 
for both the ground-state and the excited metastable states of the He-like Li ion and to photo-excitation, 
comparing and contrasting our results with previous theoretical approaches and investigate convergence issues.

We note that the eigenchannel R-matrix technique and the convergence close-coupling (CCC) method 
give similar results for electron impact ionization of H and He$^+$ \cite{bray95}. 
Owing to the success of the CCC approach a comparison of this method with the present R-matrix calculations 
permits a more conclusive evaluation of the reliability of the R-matrix method.
Therefore we expect to have similar agreement between the R-matrix with pseudo-states and 
the convergent close-coupling methods for the case of photon impact 
in the above half collision process with this He-like complex.

\begin{figure}
\begin{center}
\includegraphics[width=\textwidth]{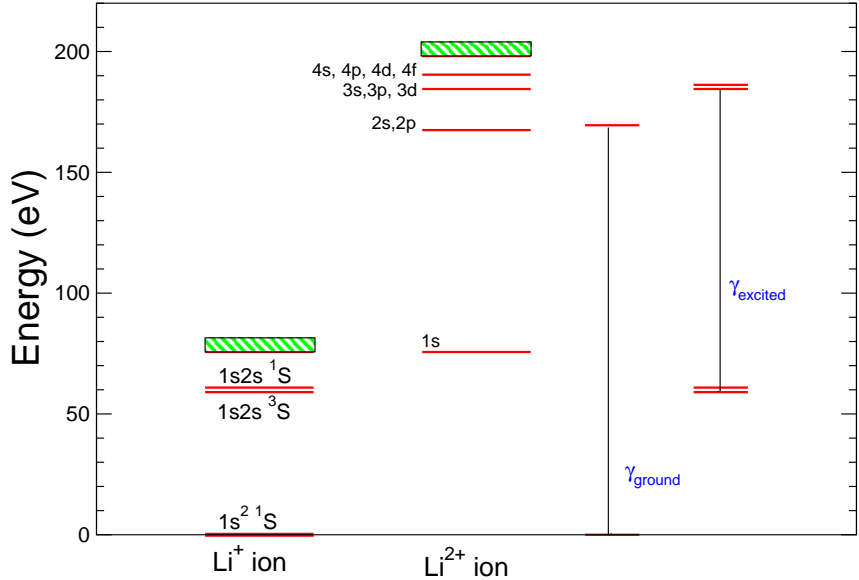}
\caption{\label{fig:overview7} (Colour online) Energy-level diagram of some of the levels of the Li$^+$ 
                                                       and Li$^{2+}$ ions. The energies are from the NIST tabulations. 
                                                     $\gamma_{\rm ground}$ and $\gamma_{\rm excited}$ are the 
                                                     lowest photon energies used to calculate the cross-sections 
                                                     for the ground state $\rm1s^2~^{1}$S and the 
                                                     $\rm 1s2s~^{1,3}$S two excited metastable states respectively. Adapted from the 
                                                     work of Kleiman and co-workers \cite{ulrich05}.}
\end{center}
\end{figure}

\begin{figure}
\begin{center}
\includegraphics[width=\textwidth]{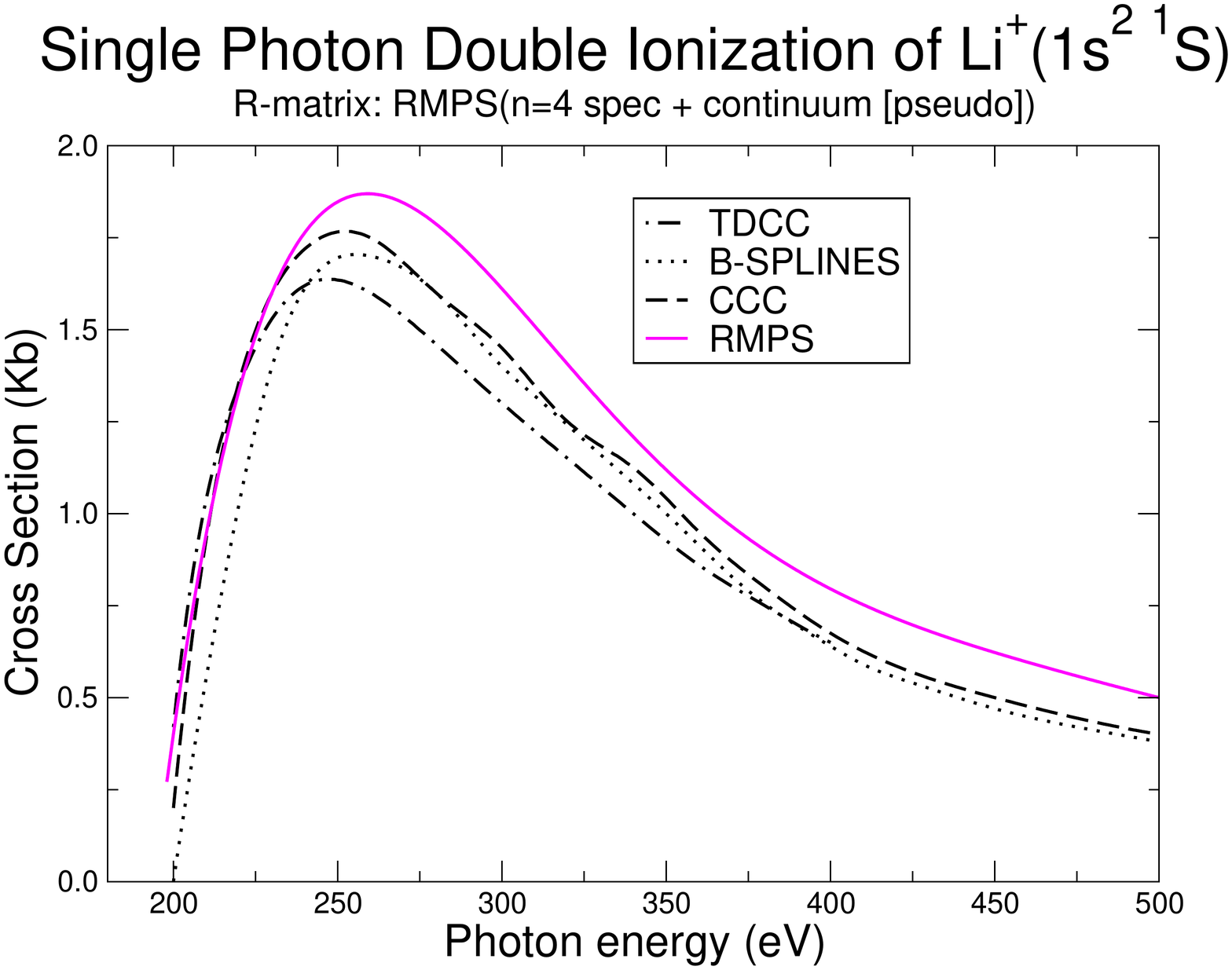}
\caption{\label{fig:overview1} (Colour online) Theoretical cross sections (Kb) for the single-photon double-ionization
                                                       of Li$^+$ ions from the ground-state for the photon energy range
                                                        200 eV to 60 eV. Results from the present R-matrix plus pseudo-states (RMPS) 
                                                        (solid line), the time-dependent close-coupling (TDCC) \cite{mitch07}
                                                          (dot-dashed line), convergent close-coupling (CCC)  \cite{bray98,bray98b,bray00,bray01}
                                                         (dashed line)  and B-splines \cite{hugo2001} (dotted line) methods are illustrated 
                                                         and included for comparison purposes \cite{McLaughlin2009}.}
\end{center}
\end{figure}

The layout of this paper is as follows. Section 2 presents a brief outline of the theoretical work. 
Section 3 details the results obtained. Section 4 presents a discussion 
and a comparison of the results obtained between various 
theoretical methods as limited experimental measurements are available. 
Finally in section 5 conclusions are drawn from the present investigation.

\section{Theory}
Photoionization cross-section calculations were performed in $LS$--coupling on
the two-electron He-like Li$\rm ^{+}$ ion using the R-matrix methodology \cite{codes,rmat,Burke2011,ICPEAC2012}. 
We use an efficient parallel version of the R-matrix codes. Details of the atomic structure employed in the present
calculations have been published before  \cite{Scully2006,Scully2007} and only a brief summary will be presented here.
 The ionization cross sections are determined by summing over all excitations above the ionization
threshold, including all single-electron excitations to the pseudo-states as well as doubly excited states. 
Single photon double ionization from the ground state has been reported on previously using this same method \cite{McLaughlin2009}. 
In the present work we use 65 and 80 levels respectively of the residual Li$\rm ^{2+}$ ion  states  using the R-matrix with
pseudo-state (RMPS) method introduced by Burke, Bartschat and co-workers \cite{Burke1969,Bartschat1996a,Bartschat1996b,Bartschat1997a,Bartschat1997b} 
and further extended by Badnell and co-workers \cite{Gorczyca1997a,Gorczyca1997b,mit99} for the close-coupling calculations.  
The basis sets consist of n=4 spectroscopic orbitals and $\overline{\rm n\ell}$ = $\overline{\rm 5\ell}$ $\dots$ $\overline{\rm 18\ell}$ 
($\ell$ =0, 1, 2, 3, and 4, i.e. $s$, $p$, $d$, $f$ and $g$ angular momentum)  
correlation/pseudo orbitals of  Li$\rm ^{2+}$ to represent the target wavefunctions. 
Basis set RMPS1 has n=4  spectroscopic orbitals and  $\overline{\rm n\ell}$ = $\overline{\rm 5\ell}$ $\dots$ $\overline{\rm 15\ell}$,  
correlation/pseudo orbitals, whereas for basis set RMPS2, we have expanded the pseudo-state representation of 
the continuum as, $\overline{\rm n\ell}$ = $\overline{\rm 5\ell}$ $\dots$ $\overline{\rm 18\ell}$,  correlation/pseudo orbitals 
along with retaining the n=4  spectroscopic orbitals in the basis.
All of these hydrogenic orbitals were determined using the AUTOSTRUCTURE
 program \cite{badnell86,badnell11} for the Li$\rm ^{2+}$ ion.
  
\begin{figure}
\begin{center}
\includegraphics[width=\textwidth]{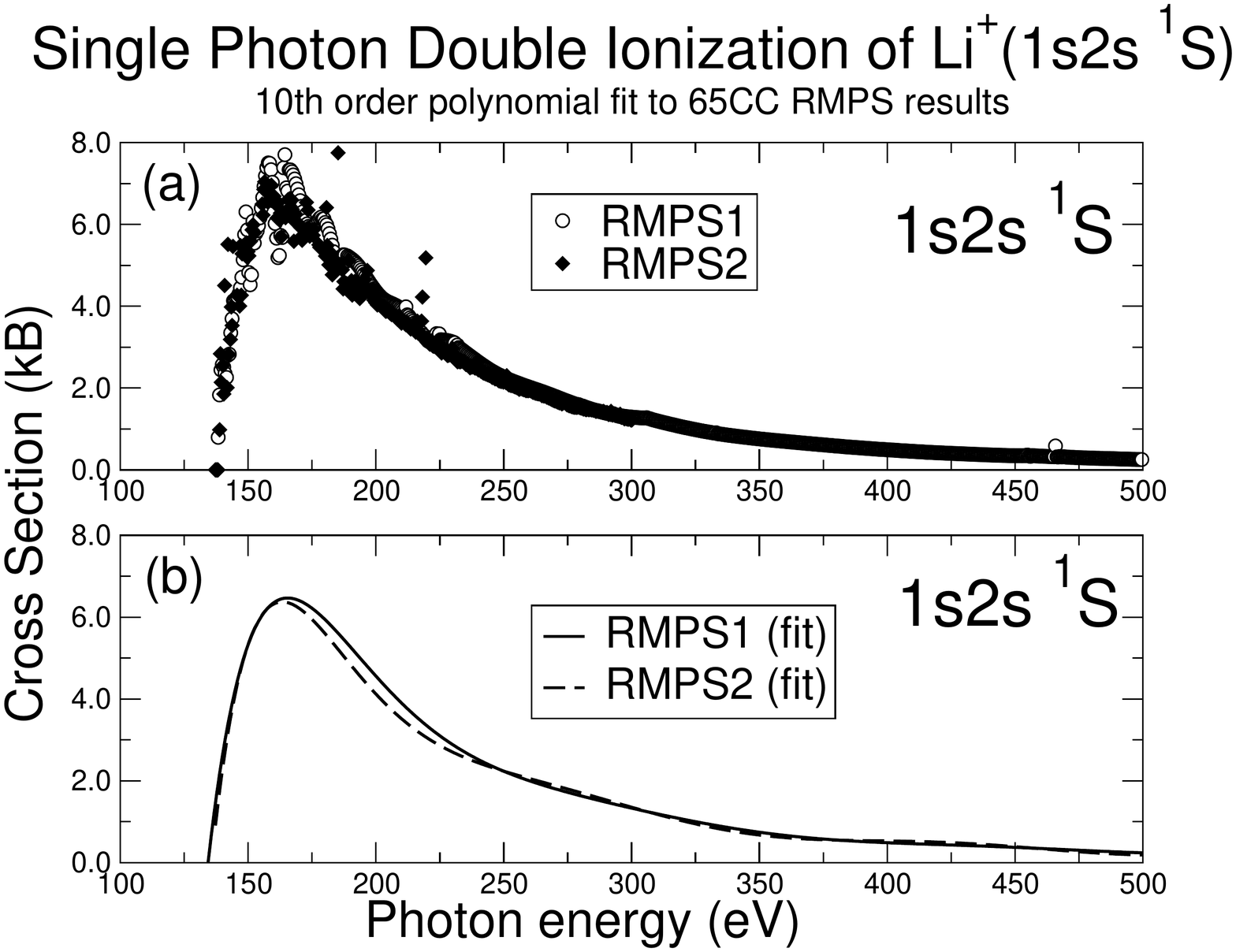}
\caption{\label{fig:overview2}   Theoretical cross sections (Kb) for the single-photon double-ionization
                                                       from the Li$^+$ (1s2s~$^1$S) metastable state for the photon energy range
                                                        100 eV to 500 eV are illustrated. (a) Raw results from the R-matrix with pseudo-states (RMPS) method
                                                        from two the different basis size (see theory section of text for explanation, open circles
                                                        basis set RMPS1, solid diamonds basis RMPS2)
                                                        are shown to illustrate the convergence. 
                                                        (b) 10$^{th}$-order polynomial  best fits to the raw data
                                                        from the 65-state model are shown, solid line, basis RMPS1, 
                                                        and dashed line, basis RMPS2.}

\end{center}
\end{figure}

For photoionization of this He-like system, 120 continuum orbitals were used and
double-electron promotions from specific base configuration sets 
described the  (Li$\rm ^{2+}$+ e$^-$) scattering wavefunction in the RMPS
calculations. In our previous work on single photon, single electron ionisation
 \cite{Scully2006,Scully2007} an energy mesh size of 13.6 $\mu$eV was required 
in order to resolve all the fine resonances in the PI cross sections for resonances lying 
below the single ionization threshold. Here since we are interested in processes 
above the ionization threshold a broader mesh size was used of 0.02 Rydbergs (272 meV).
For the excitation and ionization processes 
studied here we use models differing only in the size of the basis 
included in the close-coupling calculations, as a means of checking the convergence of our results.  
These basis sets we designate as follow;  RMPS1, in which 
we restrict the pseudo-state basis to $\overline{\rm n\ell}$ = $\overline{\rm 15\ell}$ and
RMPS2,  where we extend the pseudo-state basis to 
$\overline{\rm n\ell}$ = $\overline{\rm 18\ell}$, thus allowing for 
checks to be made on convergence of the method. For the case of a two-electron 
system and the single photon double detachment process in H$^{-}$, 
we note that a smaller pseudostate basis set (n=1 - 4 physical, 5 -  14 pseudostates,  
with $s$, $p$, $d$, $f$ and $g$ angular momentum) within the RMPS approach 
reproduced cross section results obtained using an extended pseudostate basis 
(n=1 - 4 physical, 5 -  38 pseudostates,  with $s$, $p$, $d$ and $f$ angular momentum) 
within the Intermediate Energy R-matrix Method (IERM) method \cite{Scott2012}.

\section{Results}
All the photoionization cross sections were determined in $LS$ - coupling 
with an efficient parallel version \cite{ballance06} of the R-matrix programs \cite{rmat,codes,damp}. 
Length and velocity forms of the cross sections were seen to be in excellent agreement and 
virtually indistinguishable. We have therefore chosen to plot only the length form.
 Fig. 2 shows a sample of the RMPS (using basis set RMPS1 as defined previously) 
 results for the ground-state compared to those from the time-dependent 
close-coupling (TDCC) approach \cite{ulrich05}, 
the close-coupling to the continum method (CCC) \cite{bray98,bray98b,bray00,bray01} and 
the B-spline method \cite{hugo2001}. A best fit (6$^{th}$ - order polynomial fit) to the raw results was used.
Consistent with other complexes studied such as Li, Be, Mg and Ne,
 for single-photon double-ionization \cite{Griffin09,Griffin09b, Ballance09} using the RMPS method, 
we find that the present cross-section results are larger  than those obtained from the TDCC approach.
In figure 2 it can be seen that the present RMPS results are in closer agreement 
with the convergent close-coupling (CCC) approach and the B-splines methods
at all energies  above 250 eV and merge to the CCC and B-splines results at photon impact 
energies above about 500 eV.

\begin{figure}
\begin{center}
\includegraphics[width=\textwidth]{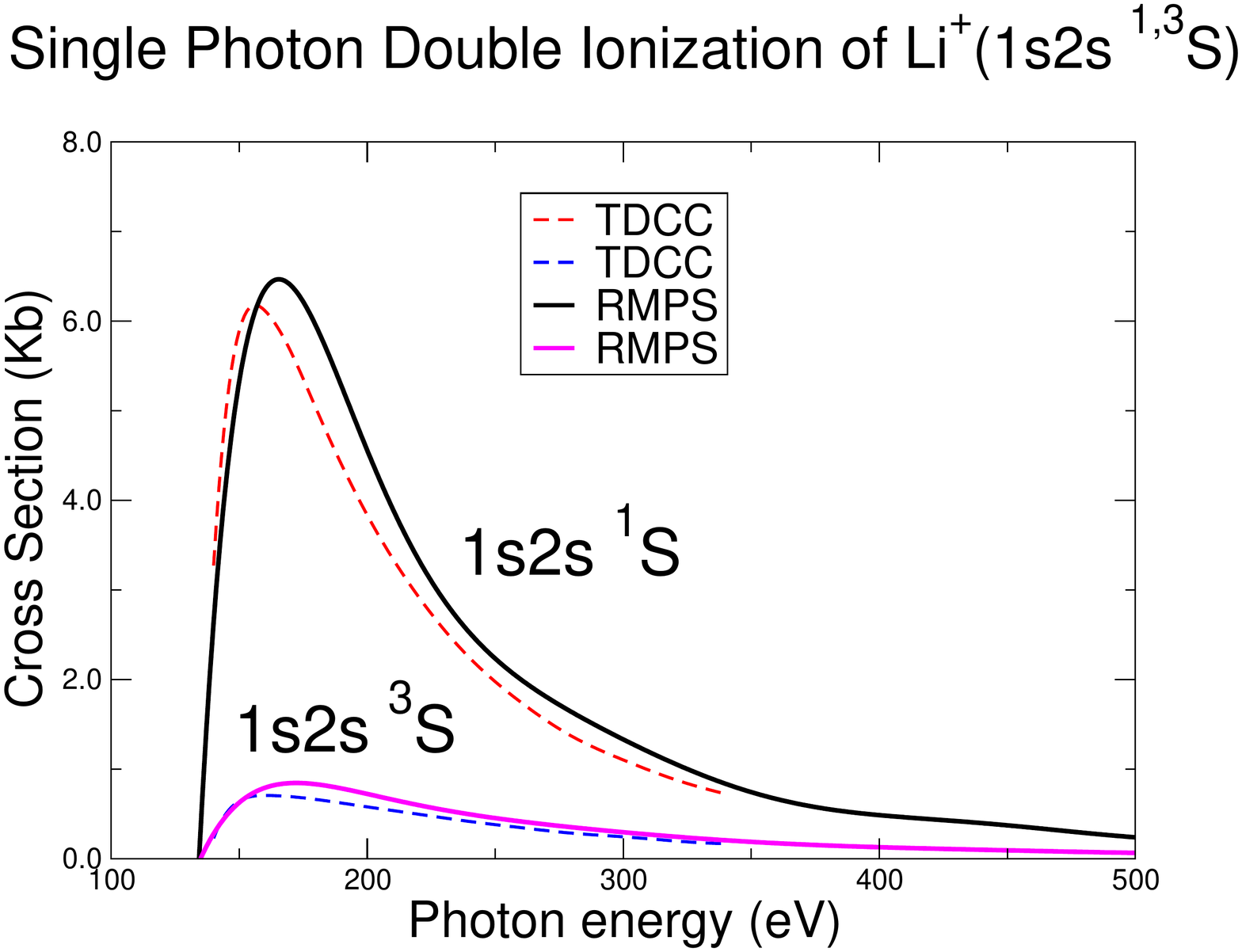}
\caption{\label{fig:overview3}  (Colour online) Theoretical cross sections (Kb) for the single-photon double-ionization
                                                       from the Li$^+$ (1s2s~$^{1,3}$S) metastable states 
                                                       for the photon energy range 100 eV to 500 eV, RMPS results (solid line)  
                                                        along with the time dependent close-coupling (TDCC) 
                                                         (dashed line) method are illustrated and included for comparison purposes.}
\end{center}
\end{figure}

To illustrate the convergence of the model employed here, we carried out double ionization calculations varying the size of the basis set used
in the close-coupling approach for the single photon double ionization from the Li$^+$ (1s2s~$^{1}$S) metastable state. 
In figures 3 (a) and 3(b), cross section results are presented from calculations using different size of basis sets, 
which are designated respectively as RMPS1 and RMPS2.  We found that results from larger basis sets gave similar 
results to those from the smaller basis. Details of the two basis sets have been outlined above.  
As clearly seen from figures 3 (a) and 3(b), the RMPS results for 
the single photon double ionization from the  Li$^+$ (1s2s~$^{1}$S) metastable state 
using the two different basis (RMPS1 and RMPS2 respectively) give comparable results. 
In figure 3 (a) there appears to be more scatter in the raw results from the larger basis (RMPS2) resulting from pseudo resonances. 
To guide the eye we present a best fit (10$^{th}$-order polynomial fit) to the raw results from the R-matrix with pseudostates
close-couplings calculations in Fig 3 (b).  For the double ionization process we conclude from our results 
presented in figures 3 (a) and 3(b)  (for all energies considered)  that the cross sections obtained 
from the larger (80 level RMPS2 calculation are adequately represented 
by those from the smaller (65 level RMPS1) model.  Subsequently in the remaining 
figures we  present only the results of our work from the 65 level model (RMPS1).

\begin{figure}
\begin{center}
\includegraphics[width=\textwidth]{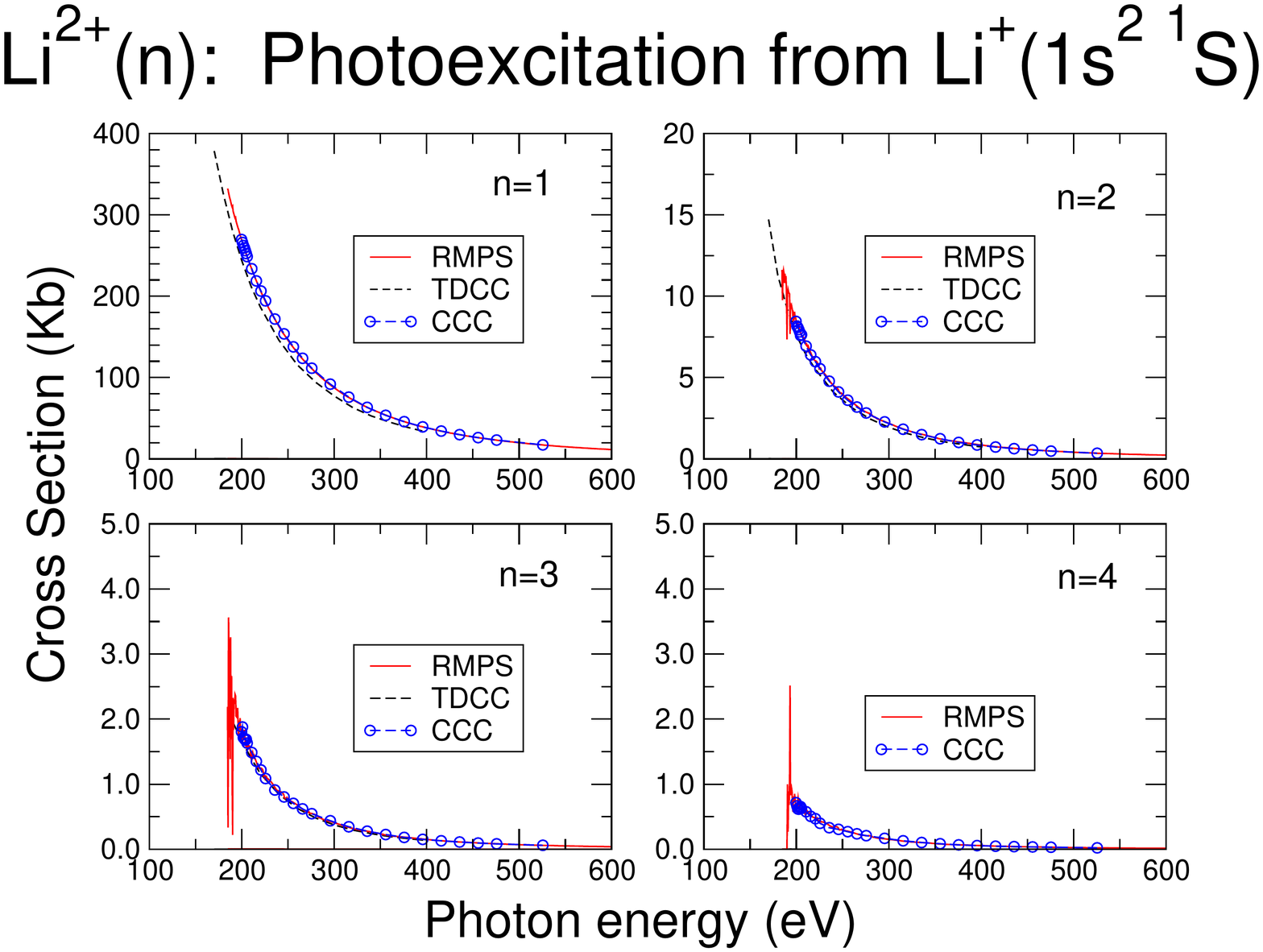}
\caption{\label{fig:overview4}  (Colour online) Theoretical cross sections (Kb) for the single-photon impact
                                                       on the Li$^+$ (1s$^2$~$^{1}$S) ground state leaving the 
                                                       residual Li$^{2+}$ ion in an excited state
                                                       for the photon energy range 100 eV to 600 eV.  
                                                       The TDCC \cite{ulrich05} and CCC \cite{bray98} results 
                                                       are included for comparison purposes.}
\end{center}
\end{figure}

\begin{figure}
\begin{center}
\includegraphics[width=\textwidth]{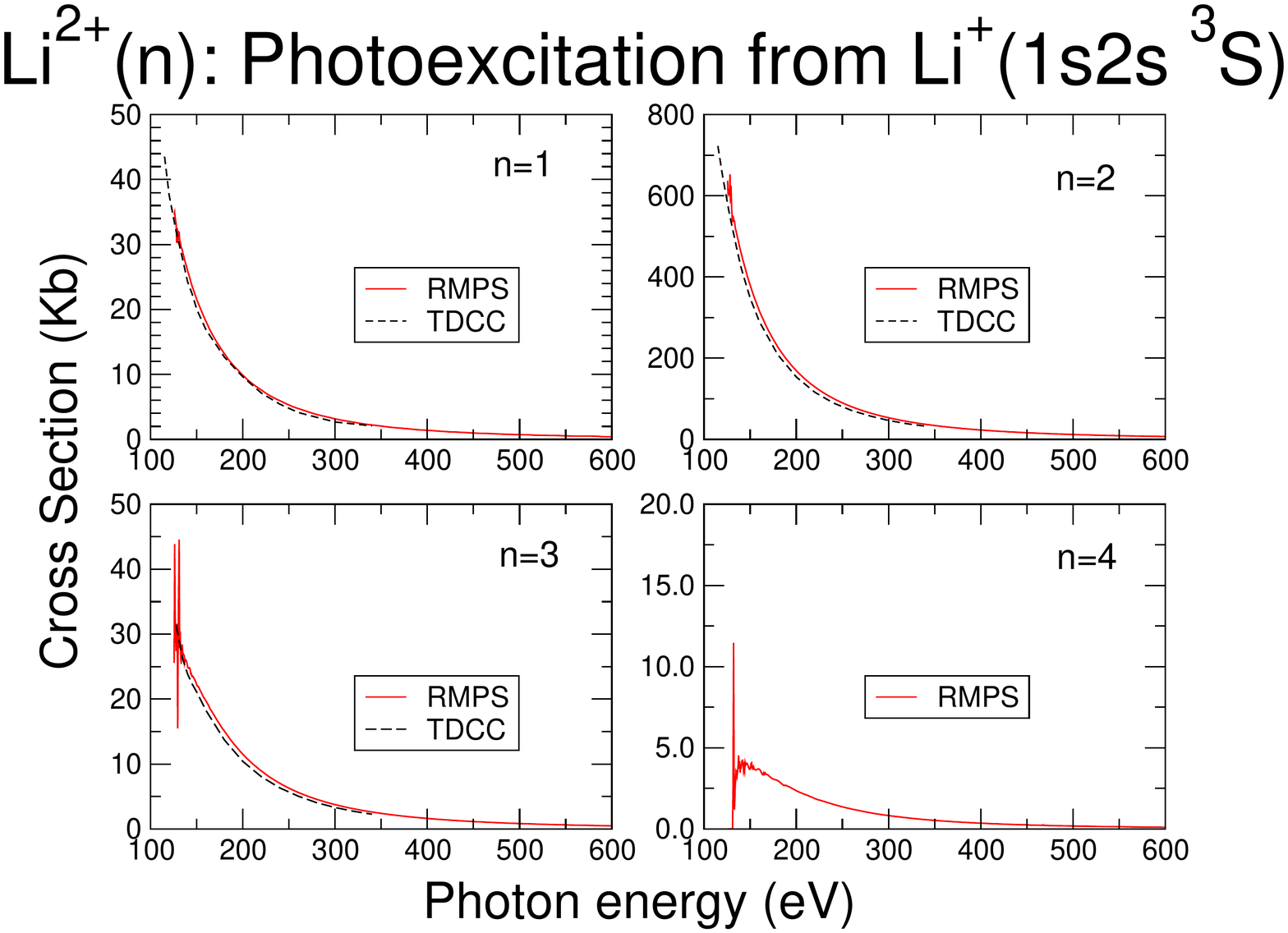}
\caption{\label{fig:overview5} (Colour online) Theoretical cross sections (Kb) for the single-photon impact
                                                       on the Li$^+$ (1s2s~$^{3}$S) metastable state leaving the 
                                                       residual Li$^{2+}$ ion in an excited state
                                                       for the photon energy range 100 eV to 600 eV.  
                                                       The TDCC results \cite{ulrich05} are included for comparison purposes.}
\end{center}
\end{figure}

\begin{figure}
\begin{center}
\includegraphics[width=\textwidth]{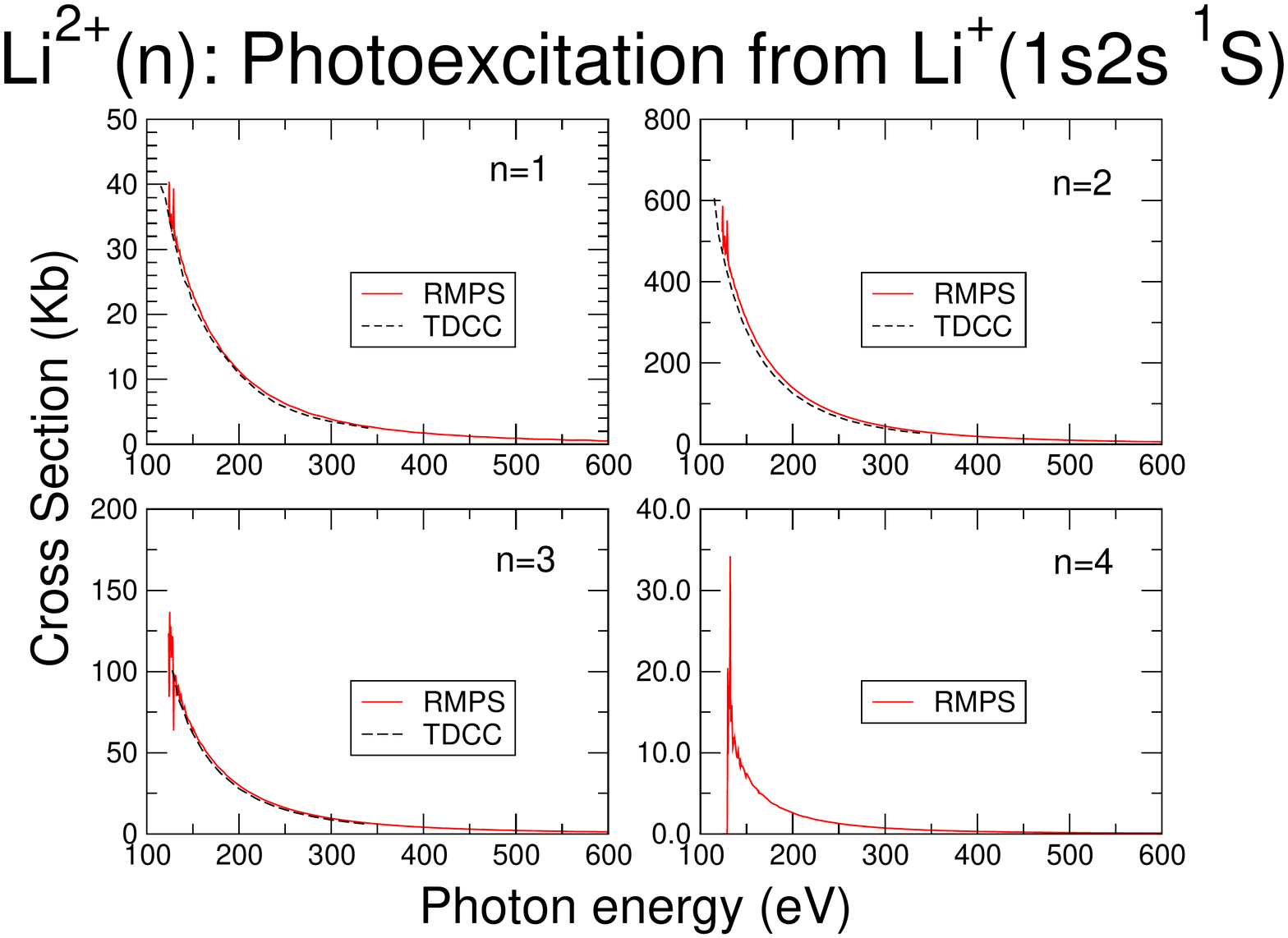}
\caption{\label{fig:overview6} (Colour online) Theoretical cross sections (Kb) for the single-photon impact
                                                       on the Li$^+$ (1s2s~$^{1}$S) metastable state leaving the 
                                                       residual Li$^{2+}$ ion in an excited state
                                                       for the photon energy range 100 eV to 600 eV.  
                                                       The TDCC results \cite{ulrich05} are included for comparison purposes.}
\end{center}
\end{figure}

\section{Discussion}
In the case of electron impact ionization of the metastable states of Li$^+$ (1s2s~$^{1,3}$S) \cite{julian07} the cross 
section from the singlet state are larger than those for the corresponding triplet metastable state. 
From the results presented in figure 4,  the situation is similar for the case of single photon impact double ionization as
 the singlet metastable state cross section is  about a factor of 6 
larger at the peak (in cross section) than that from the corresponding triplet metastable state. 
We note that cross sections obtained from both the RMPS and TDCC methods 
tend to the same value at impact energies of about 500 eV and beyond, 
 for each of the individual 1s2s~$^{1,3}$S metastable states but to different limits.
Forrey and co-workers  \cite{forrey95} have determined the different
asymptotic limits of  the ratio R (i.e., double to single photoionisation cross sections) 
for He-like ground state and the metastable  (1s2s~$^{1,3}$S) states. They have
shown that the ratio R (i.e., double to single photoionisation, expressed as a percentage) 
behaves for large Z as,
\begin{equation}
R_{Z \rightarrow \infty} \sim \frac{9}{Z^2} - \frac{3}{Z^3}
\end{equation}
for the 1s$^2$~$^{1}$S ground states,
\begin{equation}
R_{Z \rightarrow \infty} \sim   \frac{32}{Z^2} - \frac{66}{Z^3}
\end{equation}
for the 1s2s~$^{1}$S metastable states, and
\begin{equation}
R_{Z \rightarrow \infty} \sim \frac{6}{Z^2} - \frac{9}{Z^3}
\end{equation}
for the 1s2s~$^{3}$S metastable states.  
Detailed calculations by these authors, for R,  using explicitly correlated Frankowski-Perkeris-type 
functions, gave a value of 0.856 for the ground state, 1.204 for the 1s2s~$^{1}$S metastable state and 
0.304 for the corresponding 1s2s~$^{3}$S state.   The present RMPS results 
are compatible with these results. Furthermore, for He-like systems, table III of   Forrey and 
co-workers \cite{forrey95} clearly showed by explicit calculation the values of R for the 1s2s~$^{1}$S states exceed those 
of the corresponding 1s$^2$~$^{1}$S ground states for Z $\ge$ 3.

Figure 5, illustrates the present RMPS results for  photo-excitation into the n=1 - 4 excited levels 
of the residual  Li$^{2+}$ ion from the initial Li$^+$ (1s$^2$~$^{1}$S) ground-state.
Whereas, in figure 6 and 7 respectively we present the RMPS results for 
photo-excitation into the n=1 -  4 excited levels of the residual  Li$^{2+}$ 
ion from the Li$^+$ (1s2s~$^{1}$S) and Li$^+$ (1s2s~$^{3}$S)   
metastable states using the RMPS method. Figures 5, 6 and 7 
compare the present RMPS cross section results  with similar ones 
obtained from both the CCC and TDCC methods of photo-excitation of the residual 
ion for the ground and metastable states. For the ground state as can 
be seen from  the cross section results presented in figures 5, 
excitation into the ground residual level of Li$^{2+}$ (1s~$^2$S) ion dominates over all other.  
Excellent agreement between all three vastly different theoretical approaches is obtained. 
Such comparisons provide confidence in our current theoretical results.  
In the case of the Li$^+$ (1s2s~$^{1,3}$S) metastable states shown
 respectively in  figures 5 and 6, it is clearly seen that excitation into 
 the residual ion n=2 levels predominate over all other levels. 
 For these metastable states only the TDCC results \cite{ulrich05} are 
 available for comparison purposes and only for excitation 
 into the n= 1 - 3 levels of the residual  Li$^{2+}$ ion.  
From the limited data available to compare with, it is seen that the 
present RMPS results are in excellent accord with previous results  obtained 
using the TDCC method  \cite{ulrich05,mitch07,Ludlow09b,mitch10}, 
providing additional confidence in our data for applications.

\section{Conclusion}
State-of-the-art theoretical methods were used to
study the single photon double ionization of Li$^{+}$ ions 
within the R-matrix with pseudo states (RMPS) approach.  
Due to the lack of experimental data on these processes we compare the results of our study
 with those from previous theoretical studies using a variety of different methods 
in order to gauge the accuracy and quality of our work.  Given the validation with experiment 
of our previous RMPS cross section results on ground state \cite{Scully2006,Scully2007}, 
photo-absorption of Li$^+$ ions for this He-like system, and the close agreement with the 
convergent close-coupling (CCC), the time-dependent close-coupling (TDCC), 
and B-splines methods, for the ground and excited states, we expect  that the present results for single photon double ionization 
to be of comparable quality as to those for single photoionization. We hope that this current 
work might provide a stimulus for future experimental work on this complex.

\ack
BMMcL acknowledges support by the National Science Foundation through a grant to the 
Institute for Theoretical Atomic and Molecular Physics (ITAMP)
at the Harvard-Smithsonian Center for Astrophysics under the visitor's program where this work was 
completed.  ITAMP is supported by a grant from the National Science Foundation. We thank  
Dr Ulrich Kleiman for  the provision of the TDCC data and Professor Igor Bray for the CCC data. 
Helpful discussions with Professor Alex Dalgarno FRS, Dr John C Raymond and Professor Philip Stancil, 
on the astrophysical applications are gratefully acknowledged. 
Professor's Mitch S Pindzola and Connor P Ballance are gratefully acknowledged for many helpful and 
stimulating conversions on R-matrix collision theory and parallel computing throughout the course of this work.
The computational work was carried out at the National Energy Research Scientific
Computing Center in Oakland, CA, USA and on the Kraken XT5 facility at 
the National Institute for Computational Science (NICS) in Knoxville, TN, USA.
 The Kraken XT5 facility is a resource of the Extreme Science and Engineering Discovery Environment (XSEDE), 
which is supported by National Science Foundation Grant No. OCI-1053575.
%
%
%
%
\bibliographystyle{iopart-num}
\bibliography{liplus}

\end{document}